\documentclass[prd,twocolumn,showpacs,reprint,preprintnumbers,nofootinbib,amsmath,amssymb]{revtex4-2}

\RequirePackage[colorlinks=true
,urlcolor=blue
,anchorcolor=blue
,citecolor=blue
,filecolor=blue
,linkcolor=blue
,menucolor=blue
,linktocpage=true
,pdfproducer=medialab
,pdfa=true,
pdftitle={Krzysztof Jod\l{}owski - Is a covariant virtual tachyon viable?},
pdfauthor={Krzysztof Jod\l{}owski},
]{hyperref}

\usepackage{amsmath,amssymb,amsthm,amsfonts}
\usepackage{bbold}
\usepackage{graphicx,tabularx}
\usepackage{color}
\usepackage{multirow}  
\usepackage{wasysym}  
\usepackage{comment}
\usepackage{slashed}
\usepackage{enumitem}
\usepackage{cleveref}
\usepackage{rotating}
\usepackage[english]{babel}
\usepackage[justification=centerlast]{caption}
\usepackage{subcaption}
\usepackage{float}

\usepackage{breqn} 
\makeatletter 
\let\cref@old@eq@setnumber\eq@setnumber 
\def\eq@setnumber{
\cref@old@eq@setnumber
\cref@constructprefix{equation}{\cref@result}
\protected@xdef\cref@currentlabel{
[equation][\arabic{equation}][\cref@result]\p@equation\theequation}} 
\makeatother

\usepackage{cases}

\crefname{section}{Sec.}{Secs.}
\crefname{figure}{Fig.}{Figs.}
\crefname{equation}{Eq.}{Eqs.}
\crefname{appendix}{Appendix}{Appendices}
\setlist[description]{leftmargin=0.4cm}
\setlist[itemize]{leftmargin=0.4cm}

\newcommand{\be}{\begin{equation}\begin{aligned}}
\newcommand{\ee}{\end{aligned}\end{equation}}

\newcommand{\beq}{\begin{equation}}
\newcommand{\eeq}{\end{equation}}
\newcommand{\beqa}{\begin{eqnarray}}
\newcommand{\eeqa}{\end{eqnarray}}

\renewcommand{\eqref}[1]{Eq.~(\ref{#1})}

\newcommand{\eg}{e.g.}
\newcommand{\ie}{i.e.}

\RequirePackage[normalem]{ulem}

\DeclareUnicodeCharacter{2212}{\textendash}

\makeatletter
\def\l@subsubsection#1#2{}
\makeatother

\usepackage{multirow}
\usepackage{makecell}

\usepackage{fontawesome}

\usepackage[T1]{fontenc}

\newcommand{\textdaggerrotccw}{\raisebox{-0.0ex}{\rotatebox{90}{\textdagger}}}
\newcommand{\textdaggerrotccww}{\raisebox{1.0ex}{\rotatebox{-90}{\textdagger}}}

\newcommand{\TODO}[1]{\textcolor{green}{TODO}}
\RequirePackage[normalem]{ulem}
 
\begin{document}

\title{Is a covariant virtual tachyon viable?}

\author{Krzysztof Jod\l{}owski}
\email{krzysztofjjodlowski@gmail.com}

\affiliation{Department of Physics and Institute of Theoretical Physics \\Nanjing Normal University\char`,{} Nanjing\char`,{} 210023\char`,{} China}
\affiliation{Particle Theory and Cosmology Group\char`,{} Center for Theoretical Physics of the Universe\char`,{} Institute for Basic Science (IBS)\char`,{} Daejeon\char`,{} 34126\char`,{} Korea}

\begin{abstract}
    Sidney Coleman has noted that superluminal particles or observers would be able to go back in time and have no definite trajectory according to subluminal observers, while not violating Lorentz invariance~\cite{Griffiths_Derbes_Sohn_2022}. 
    Recently, Dragan and Ekert have significantly developed similar ideas even further, which lead to formulation of ``quantum principle of  relativity'' that intimately links the two theories~\cite{Dragan:2019grn}.
    However, field theory descriptions of an on-shell tachyon, described by scalar field $\phi$ with negative mass squared parameter, lead to violation of basic principles of relativity or quantum mechanics.
    In this work, we investigate whether purely virtual tachyons can be consistent within the fakeon framework—the only known viable formulation of purely virtual particles. 
    We show that the supports of the real part of the tachyon Feynman propagator, which is also the principal-value propagator (PVP), and the Wheeler propagator intersect only on the light cone.
    We show that PVP cannot be time-ordered two-point function of a local Hermitian field, excluding the direct extension of the canonical formalism to free fakeon tachyon.
    Finally, with an imposed tachyonic PVP, interactions with electrons generate composition-dependent accelerations that violate the weak equivalence principle, and we provide a quantitative bound on this scenario.
    Our analysis identifies obstacles to formulating covariant quantum field theory of interacting virtual tachyon.
    \begin{center}
        \textdaggerrotccww\,
                \textit{Dedicated to my Grandma Urszula on occasion of 96th birthday}
        \,\textdaggerrotccw
    \end{center}
\end{abstract}

\maketitle

\section{Introduction}
\label{sec:introduction}
While many authors considered superluminal observers or reference frame transformations, such objects are plagued by mathematical inconsistencies, physical paradoxes~\cite{Tolman1917,10.1119/1.1986200}, or they break one or more basic laws of physics; for brief overview, see~\cite{Kamoi:1971qb,Paczos:2023mof}.
On the other hand, their peculiar behavior has fascinated not only science-fiction writers, but also prompted serious academic works, starting with Bilaniuk et al.~\cite{Bilaniuk:1962zz}.

While the question if superluminal objects are physically consistent in the framework of classical special relativity is difficult to settle, as evidenced by comments by, \eg, DeWitt~\cite{Bilaniuk:1969vd} and Thouless~\cite{Thouless:1969ut}—who noted that while real (on-shell) tachyons present serious challenges for theory consistency, virtual (off-shell) tachyons significantly relax these constraints—the analysis is even more murky when one accounts for fundamental indeterminism of the quantum world~\cite{Dragan:2019grn}.

In fact, Ref.~\cite{Dragan:2019grn} prompted recent resurgence of interest in tachyons, since it has made several observations directly linking non-causal aspects of physics of superluminal observers and fundamentals of quantum mechanics (QM).
If correct, that would mean the latter theory can be understood as largely originating from more basic and intuitive principle - the (extended) principle of relativity, precisely formulated in~\cite{Dragan:2019grn} as ``quantum principle of relativity''.
While the idea of explaining the origin of quantum superpositions in such a way was critically examined within special relativity and non-relativistic QM in~\cite{Grudka:2023mrp}—see also comments examining~\cite{Dragan:2019grn}  along different directions~\cite{DelSanto:2022oku,Grudka:2021fdq,Horodecki:2023hvf} together with corresponding responses~\cite{Dragan_2022,Dragan_2023,Dragan:2023kpa}—the ultimate test of such program is construction of a viable quantum field theory (QFT) of tachyons, defined as scalars with negative mass squared parameter.

Motivation for considering QFT of tachyons is also provided by the fact that they are predicted by asymptotic freedom scenario in higher-derivative quantum gravity~\cite{Avramidi:1985ki}.
Tachyons are also invoked in constructions of interesting Carrollian (which corresponds to Poincare group contraction in the limit $c\to 0$~\cite{Levy-Leblond:1965dsc,SenGupta:1966qer}) models by starting with tachyonic matter in a Lorentzian theory~\cite{Najafizadeh:2024imn,deBoer:2023fnj,Ecker:2024czx}.
Finally, we investigate whether tachyons could be purely virtual particles, called fakeons~\cite{Anselmi:2017yux,Anselmi:2018kgz,Anselmi:2018tmf}, and establish an obstruction to the direct extension of their canonical free-field realization. 

We would also like to point out  that purely virtual tachyons, by construction, evade the most serious problems of such theories - namely that, since the little group is not compact, the phase space of tachyons is unbounded leading to purely kinematic divergences~\cite{Mrowczynski1983}.
Moreover, formulation of scattering theory for on-shell tachyon is problematic, there are divergences in elastic scattering of SM particles mediated by virtual tachyon, and optical theorem does not hold (unitarity is violated) with tachyonic Feynman propagator (FP)~\cite{Jodlowski:2024rut}.
Therefore, purely virtual tachyon field deserves further investigation - its origin from covariant quantization as well as formulation of interactions with Standard Model (SM) fields.
For the first time, we show that, under the principal-value propagator (PVP) prescription, \eg, electron-tachyon interaction generates matter-dependent scalar backgrounds and composition-dependent accelerations violating the weak equivalence principle.

Before proceeding further, let us make several remarks.
First, as recently discussed in detail in Ref.~\cite{Kip:2024uha}, the Higgs field can be viewed as self-interacting tachyon,\footnote{The bare mass squared is not directly physical quantity~\cite{Weinberg_1995,Coleman:2011xi,Srednicki:2007qs}. In particular, it is not an observable, which differs from tachyon that, by definition, has physical negative squared mass.} that can develop either vanishing (proper tachyon) or non-zero vacuum expectation value (the option realized within SM) through tadpole diagrams.
After Dyson series resummation of the Higgs propagator, the physical mass squared is positive—such purely diagrammatic approach does not invoke concepts like spontaneous symmetry breaking, point emphasized in~\cite{Kip:2024uha}.
Therefore, our work can be seen as exploring the vanishing tadpole  solution, which was promptly dismissed in~\cite{Kip:2024uha} due to tachyon problems with causality; we also note that the Higgs mechanism was brought up as one of main motivations for investigating tachyons in~\cite{Dragan2023,Paczos:2023mof}.
Therefore, we investigate whether purely virtual prescriptions avoid these problems for perturbations about the unstable configuration.

Second, already DeWitt~\cite{Bilaniuk:1969vd} and Thouless~\cite{Thouless:1969ut} have noted, working exclusively within  special relativistic setup, that off-shell tachyons \textit{might} be compatible with known laws of physics, although no clear conclusion was drawn by these authors.
Within QFT, several works~\cite{Murphy:1972ii,Jue:1973mr} have considered tachyon field representing a virtual state, basing it on the claim that tachyon commutator identically vanishes.
We believe that such an approach is not promising, since it  does not correspond to the standard properties of the commutator, which is given by \cref{eq:PJ}.
Moreover, an identically vanishing commutator cannot reproduce the standard canonical retarded and advanced Green's functions of the tachyonic Klein--Gordon operator.
In our approach, the commutator vanishes at spacelike separation, and the advanced and retarded Green's functions have support in the backward and forward lightcones, respectively.
On the other hand, tachyonic commutator (and therefore so do advanced and retarded Green's functions) grows exponentially in time-like direction, illustrating unstable nature of the tachyon.

Third, consistent formulation of purely virtual particles for positive mass squared - which is motivated, \eg, by quantization of quantum gravity theories, \eg, renormalizable quadratic gravity~\cite{Stelle:1976gc}, which possess ghosts - has been achieved only recently within the fakeon framework~\cite{Anselmi:2017yux,Anselmi:2018kgz,Anselmi:2018tmf,Piva:2023bcf}.
It yields a Lorentz-invariant and perturbatively unitary physical $S$ matrix,  obtained after the fakeon projection.
The cost is abandoning analyticity of the probability amplitudes across thresholds involving fakeons. In the Euclidean formulation, the continuation across such a threshold is defined by averaging the two analytic continuations that bypass it on opposite sides~\cite{Anselmi:2025uzj}.
While the currently known proofs of mathematical consistency of the fakeon projection require $m^2>0$, it is \textit{not} known if some generalized prescription exists or not~\cite{Anselmi_Piva}; also see~\cite{Anselmi:2025uda} for discussion of tachyonic fakeon in a different context.

Finally, tachyons are also studied in the recent bloom in exploration Carrollian physics.
Indeed, since in such a setting the speed of light vanishes, no movement for ordinary matter is allowed, and any non-zero velocity is necessarily of superluminal origin, which leads to  non-trivial constructions~\cite{Najafizadeh:2024imn,deBoer:2023fnj,Ecker:2024czx}.

We work in the natural units, $\hbar=c=1$, and Minkowski space with mostly minus metric.

\section{Tachyonic Green's functions}
\label{sec:Green_functions}
We will consider the following Lagrangian, extending the Standard Model (SM) by kinetic, mass term, and Yukawa interaction of tachyon with a Standard Model (SM) massive fermion $\psi$,
\be
   \mathcal{L}=  \mathcal{L}_{\mathrm{SM}} + \frac12 \partial_\mu \phi \, \partial^\mu \phi - \frac12  m_\phi^2 \,\phi^2 + g \,\phi  \,\bar{\psi}\psi
\,,
\label{eq:Lag_KG_tach}
\ee
where $m_\phi^2<0$, and $g$ is a dimensionless coupling constant.
For concreteness, we will assume further that $\psi$ represents an electron, and treat this Yukawa interaction as a low-energy effective coupling in the electroweak broken phase.
Since tachyon corresponds to small perturbations around the  unstable, $\phi\sim 0$, saddle point of the potential, by construction we neglect the impact of $\phi$ self-interactions, in particular the (possibly radiatively generated) quartic term.

Let us discuss the free theory first.
The equations of motion resulting from \cref{eq:Lag_KG_tach} are 
\be
    \left(\partial_\mu \partial^\mu + m_\phi^2\right) \phi(x) = 0 
    \,.
\label{eq:KG}
\ee
We begin by discussing the LI solutions to \cref{eq:KG} as well as LI Green's functions in position space.
We provide complete expression for tachyonic Feynman propagator (also known as the Dhar-Sudarshan propagator~\cite{Dhar:1968hkz}) and the Wheeler propagator in position space.

We also provide connection of the LI solution and Green's functions to the quantum field $\chi(x)$ describing the tachyon.
Here, we use the field decomposition and commutation relations specified in \cref{sec:fakeon}, where more of its properties are discussed.

\begin{figure}[tb]
    \includegraphics[scale=0.18]{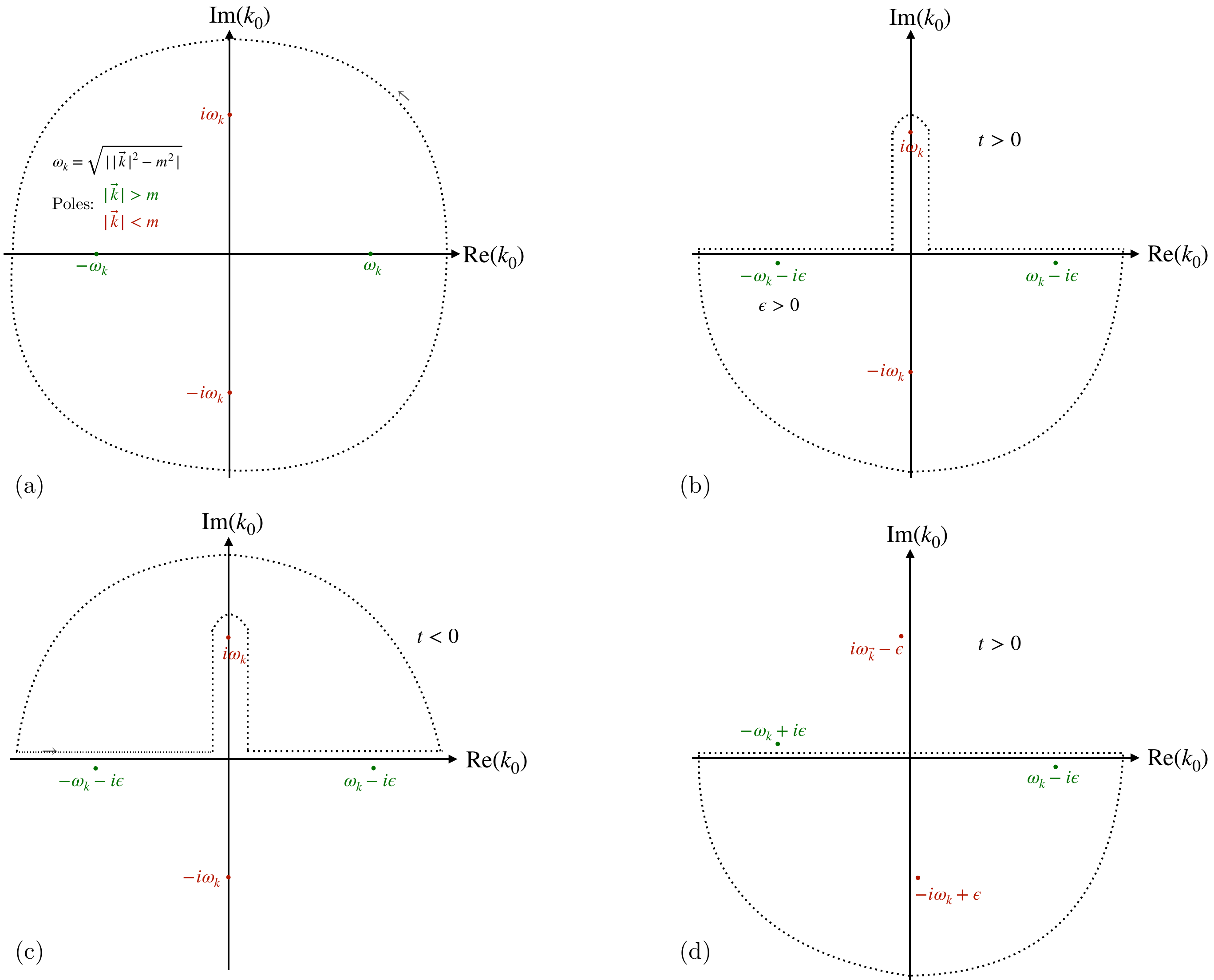}
    \caption{
            Integration contours in the complex $k^0$ plane for: (a) the commutator function; (b) and (c) the retarded Green's function for $t>0$ and $t<0$, respectively; and (d) the Feynman propagator for $t>0$.
        }
    \label{fig:contours}
\end{figure}

We will use the contours shown in \cref{fig:contours}.
These are the same as in the standard case, except for the deformation in the panels (b) and (c) that are used for computation of the retarded Green's function.
We have denoted the real and imaginary poles in green and red, respectively.
The tachyon energy is $E_k=\pm\omega_k$ for $|\vec{k}|>|m_\phi|$, while $E_k=\pm i\Omega_k$ for $|\vec{k}|<|m_\phi|$, where $\omega_k=\sqrt{|\vec{k}|^2-|m_\phi|^2}$ and $\Omega_k=\sqrt{|m_\phi|^2-|\vec{k}|^2}$.

\paragraph{Commutator function}
The vacuum expectation value of the commutator of the field $\chi$ given by \cref{eq:tach_psi} is
\be
\label{eq:Com}
    &\langle 0 | [\chi(x), \chi(y)] | 0 \rangle  = \langle 0 | [\phi(x), \phi^\dagger(y)] | 0 \rangle  \\
    &=\langle 0 | \phi(x) \phi^\dagger(y) | 0 \rangle = i\Delta(x-y)
    \,,
\ee
where $i\,\Delta(x)$ is the (Pauli-Jordan) commutator function. We used $\phi(x)|0\rangle=0$ for all $x$ and the antisymmetry of $\langle0|\phi(x)\phi^\dagger(y)|0\rangle$, which follows from the commutation relations in \cref{sec:fakeon}.

Using the contour shown in panel (a) in \cref{fig:contours}, one calculates $i\,\Delta(x)$ by straightforward computation analogous (discussed in detail in Sec. 2.3 of Ref.~\cite{Scharf:1996zi}) to the bradyon case,
\begin{align}
\label{eq:PJ}
    &i\Delta(x-y) = \int_{C_{(a)}} \frac{d^4 k}{(2\pi)^4} \, \frac{-i\,  e^{-ik(x-y)}}{k^2-m_\phi^2}  \nonumber\\
    &= -i\int \frac{d^3\vec{k}}{(2\pi)^3\, E_k }\, e^{-i\vec{k}\cdot(\vec{x}-\vec{y})}  \sin(E_k (x^0-y^0)) \nonumber\\
    & = -i\int_{|\vec{k}|>|m_\phi|} \frac{d^3\vec{k}}{(2\pi)^3\, \omega_k }\, e^{-i\vec{k}\cdot(\vec{x}-\vec{y})} \sin(\omega_k (x^0-y^0))  \nonumber\\
    &   -i \int_{|\vec{k}|<|m_\phi|} \frac{d^3\vec{k}}{(2\pi)^3\, i\Omega_k }\, e^{-i\vec{k}\cdot(\vec{x}-\vec{y})} \sin(i\Omega_k (x^0-y^0)) \nonumber\\
    &=\frac{-i\, \mathrm{sgn}(x^0-y^0)}{2\pi} \Big[\delta((x-y)^2) \nonumber\\
    &+ \theta((x-y)^2) \, \frac{|m_\phi|}{2\sqrt{(x-y)^2}} I_1(|m_\phi|\sqrt{(x-y)^2}) \Big] \,,
\end{align}
where $I_1(x)$ is the modified Bessel function of the first kind and $\theta(x)$ is the Heaviside step function.
This matches the expressions stated in~\cite{Schroer:1971pa,Streit:1971qa}.
It is worth pointing out that $\Delta(x-y)$ can be obtained from  $\Delta_{\mathrm{bradyon}}(x-y)$ by analytic continuation, $m\to \pm i\, m$.

The desirable properties of the commutator are: \\
\noindent $\Delta(x-y)$ is local, LI solution of \cref{eq:KG}, and vanishes outside the lightcone, so the field $\chi(x)$ satisfies microcausality. 
Moreover, $\Delta(x-y)$ satisfies the initial conditions for \cref{eq:KG},
\be
    \Delta(0,\vec{x})=0\,, \quad 
    \partial_{t}\,\Delta(0,\vec{x})=-\delta^3(\vec{x})
    \,, 
\ee
which translates into satisfying the equal time canonical commutation relation (CCR),
\be 
\label{eq:CCR_symplectic}
    [\chi(0,\vec{x}), \partial_{t}\chi(0,\vec{y})]=i\delta^3(\vec{x}-\vec{y})
\,. 
\ee
Since $\Delta$ is LI, so is \cref{eq:CCR_symplectic}; this can also be seen from the fact that the (classical) symplectic form for KG operator is LI for both signs of $m^2$ term.

However, the asymptotic behavior of the commutator in the time-like regime is problematic, since it grows exponentially~\cite{10.1143/PTP.24.171}.
Indeed, for large proper time $\tau = \sqrt{x^2}\gg 1/|m_\phi|$,
\be
    i\Delta(x) \sim  \frac{-i\,\text{sgn}(x^0)}{4\pi}\sqrt{\frac{|m_\phi|}{2\pi (x^2)^{3/2}}}\; e^{|m_\phi| \sqrt{x^2}} 
\,.
\ee
The instability has an obvious physical interpretation -- generic long-wavelength perturbations grow away from the unstable configuration. If self-interactions stabilize the potential, this can drive the field toward a different minimum.
This well-known fact provides the first check that our covariant formalism is on the right track—since we want to understand tachyons, we must work around the $\chi=0$ configuration, and the resulting  commutator of such field indeed indicates presence of instability.

\paragraph{Advanced/retarded Green's functions}
For bradyons, the advanced/retarded Green's functions play key role in solving the initial value problem, and are given by
\be
    \Delta_A(x)= -\theta(-x^0)\, \Delta(x) \,, \quad
    \Delta_R(x)= \theta(x^0)\, \Delta(x)
\,.
\ee

The same relation holds for the tachyonic analogues.
The difference is that  for the retarded Green's functions, we need to deform the contour, as shown in panels (b) and (c), since only such contour (and homologically equivalent ones) guarantees satisfying causality, $\Delta_R(x^0<0)=0$.

The resulting formulas are as follows:
\be
\label{eq:Delta_Adv_ret}
    &\Delta_R(x)= \Delta_A(-x) = \theta(x^0)\, \Delta(x)  \\ 
    &=\frac{-\theta(x^0)}{2\pi} \left(\delta(x^2) +\theta(x^2) \, \frac{|m_\phi|}{2\sqrt{x^2}} I_1(|m_\phi|\sqrt{x^2}) \right) 
\,.
\ee
Therefore, $\Delta_R(x)$ grows exponentially in the future timelike direction, while $\Delta_A(x)$-in the past.
Consequently, generic long-wavelength perturbations grow exponentially under retarded initial-value evolution.

\paragraph{Feynman propagator}
To compute the FP, one uses the contour shown in panel (d) in \cref{fig:contours} for $t>0$, and its complement for $t<0$.
The result is
\be
\label{eq:FP}
    &\Delta_{F}(x) =  \int \frac{d^4 k}{(2\pi)^4} \, \frac{e^{-ik x}}{k^2-m_\phi^2+i\epsilon} \\
    &=-\frac{\delta(x^2)}{4\pi} +i\frac{\theta(x^2)\, |m_\phi|}{4\pi^2 \sqrt{x^2}} K_1(|m_\phi|\sqrt{x^2})
    +\frac{\theta(-x^2)\,|m_\phi|}{8\pi \sqrt{-x^2}} \times \\
    & \times 
    \left[ J_1(|m_\phi|\sqrt{-x^2}) + i Y_1(|m_\phi|\sqrt{-x^2}) \right]   
    \,.
\ee
Similarly to the commutator, tachyonic FP can be obtained from the bradyon case by analytic continuation in mass, $m\to -i\,m$, where~\cref{eq:FP} is understood distributionally.

On the other hand, the causal behavior of bradyon and tachyon FP are different.
For $x^2>0$, the tachyon FP is purely imaginary and decays exponentially,
\be
    \Delta_F(x) \sim i \frac{\sqrt{|m_\phi|}}{4\sqrt{2}\,\pi^{3/2} (x^2)^{3/4}} \,e^{-|m_\phi|\sqrt{x^2}}
\,,
\ee
while for $x^2<0$, the FP is oscillating,
\be
    \Delta_F(x) \sim \frac{\sqrt{|m_\phi|}(-1-i)}{8\pi^{3/2} (-x^2)^{3/4}} \,e^{i|m_\phi|\sqrt{-x^2}}
    \,.
\ee
In particular, the real part of FP has support on and \textit{outside} the lightcone, and at large spacelike separation its oscillatory factor is $\cos(\pi/4 + |m_\phi|\sqrt{-x^2})$.
In \cref{sec:nonrel_potential}, we discuss in detail physical consequences of such behavior.

\paragraph{Wheeler propagator}
The arithmetic average of advanced and retarded Green's functions is  called the Wheeler propagator~\cite{Bollini:1998hj,Koksma:2010zy},
\be
\label{eq:Delta_W}
    \Delta_W(x) = \frac 12 \left( \Delta_R(x) + \Delta_A(x) \right)  = \frac 12\, \mathrm{sgn}(x^0)\, \Delta(x)
\,,
\ee
because it was notably used in Wheeler-Feynman (time-symmetric) electrodynamics~\cite{Wheeler:1945ps,Wheeler:1949hn}, which can also be viewed as relativistic action at a distance between a source and an absorber; it was also used within general relativity as a computational device~\cite{Damour:2001bu}.

For tachyons, the Wheeler propagator in position space is
\be
\label{eq:Delta_W_tach}
    \Delta_W(x) 
    &=\frac{-1}{4\pi} \left(\delta(x^2) + \theta(x^2) \, \frac{|m_\phi|}{2\sqrt{x^2}} I_1(|m_\phi|\sqrt{x^2}) \right) 
\,.
\ee
Therefore, it has support on and inside the lightcone, and it blows up exponentially for both past and future in the timelike regime.

For fields with $m^2\geqslant 0$, the Wheeler propagator overlaps with the PVP, $\bar{\Delta}(x)$, which is defined as the arithmetic average of Feynman and Dyson (also called anti-Feynman) propagators,
\be
\label{eq:Delta_PVP}
    \bar{\Delta}(x) =   \frac 12 \left( \Delta_F(x) + \Delta_{D}(x) \right)
\,.
\ee
This equation follows from the Sochocki--Plemelj formula, 
\be
    \lim_{\epsilon \to 0^+} \frac 1{x \pm i\epsilon} = \mp i\pi \delta(x) + \mathrm{PV}\left(\frac 1x \right)  \,,
\label{eq:Soch}
\ee
applied to $x=k^2- m^2$,
\begin{align}
\label{eq:PV}
    & \lim_{\epsilon \to 0^+}  \frac12 \left[ \frac 1{k^2 - m^2+i\epsilon} + \frac 1{k^2 - m^2-i\epsilon} \right]  \nonumber\\
    &    = \lim_{\epsilon \to 0^+}  \frac{k^2 - m^2}{(k^2 - m^2)^2 +\epsilon^2}  = \mathrm{PV}\left(\frac 1{k^2 - m^2} \right) 
\,.
\end{align}
Clearly, the PVP is time-symmetric and it equals the real part of $\Delta_F(x)$.
Moreover, since for bradyons there are no imaginary poles, 
\be
\label{eq:PV_W_equality}
    \bar{\Delta}(x)=\Delta_W(x)
\,.
\ee
The last relation does \textit{not} hold for tachyons, resulting in important physical consequences for viability of the covariant virtual tachyon theory. 
\textit{After} completion of this work, we have learned from Jan Dereziński that $\bar{\Delta}(x) \neq \Delta_W(x)$ for tachyons has already been noted in~\cite{Derezinski:2024cur}, but without analysis of its physical consequences.
Moreover, we have realized that a tachyonic PVP \textit{cannot} be obtained as a time-ordered two-point correlation function of a local Hermitian field satisfying CCR given by \cref{eq:CCR_symplectic}.
For example, at spacelike separation, $\chi$ leads to a real two-point correlation function, whereas $i\bar\Delta$ is imaginary and nonzero, showing the claimed impossibility.
(This argument is independent of the indefinite metric.)

The two functions entering \cref{eq:PV_W_equality} have very different behavior: $\Delta_W(x)$ has support on and inside the lightcone, while $\bar{\Delta}(x)$ has support on and outside it.
If virtual tachyon is described by the PVP, the resulting effective matter equations depend on spacelike-separated data. 
Contrary to fakeons, the tail is \textit{spacelike}, extending to arbitrarily large invariant distances.
Moreover, since the Wheeler-Feynman absorber mechanism relies on lightlike  propagation, it also cannot be directly transferred to tachyonic PVP.

To briefly summarize the main result of this section—-we have provided complete expressions for the tachyonic LI commutator and Green's functions in position space.
Crucially, we have shown that the Wheeler and real part of the FP differ away from the lightcone for tachyon, and that the latter cannot be obtained as time-ordered two-point correlation function of a local Hermitian field with the canonical commutator.

\section{Canonical quantization of a free tachyon fakeon}
\label{sec:fakeon}
Purely virtual fields have been consistently constructed within the fakeon framework~\cite{Anselmi:2017yux,Anselmi:2018kgz,Anselmi:2018tmf}. For positive mass squared, the tree-level prescription reduces to the PVP, while loop amplitudes require the full fakeon prescription, including average continuation across thresholds involving fakeons.
In particular, relativistic covariance and perturbative unitarity of the $S$ matrix after projecting out the virtual field was proved~\cite{Anselmi:2021hab}, which was applied to ghosts, plagued by inconsistencies within standard formalism.

In the canonical QFT formalism, describing fakeons requires working with a Hermitian field composed of a \textit{pair} of particle-ghost—see Sec. 3.4 in~\cite{Anselmi:2022toe}.
Since our goal is showing that constructing covariant virtual field for tachyon is impossible within canonical formalism, we adapt ansatz inspired by the fakeon decomposition; notabene, it is particularly suitable to handle the unstable, $|\vec{k}|<|m_\phi|$ modes.
Within such framework, we show that the tachyon field $\chi(x)$ satisfies microcausality and equations of motion, as well as that its commutator and Green's functions are LI.

Let us consider the following tachyon field:
\be
    \sqrt{2}\,\chi(x)= \phi(x) + \phi^\dagger(x)
    \,,
\label{eq:tach_psi}
\ee
where the non-Hermitian field $\phi(x)$ contains only annihilation operators, 
\be
    &\phi(x) = \phi_>(x)+\phi_<(x) \\
    &= \int_{|\vec{k}|>|m_\phi|} \frac{d^3\vec{k}}{2\,\omega_k\, (2\pi)^3} \Big[ a_{+\vec{k}}\,\, e^{-i(\omega_k x^0 - \vec{k}\cdot\vec{x})}  \\
    &\quad\quad\quad\quad\quad\quad\quad\quad\quad\quad\quad\quad\quad\quad + a_{-\vec{k}}\,\, e^{i(\omega_k x^0 - \vec{k}\cdot\vec{x})} \Big]   \\
    &+  \int_{|\vec{k}|<|m_\phi|} \frac{d^3\vec{k}}{2\,\Omega_k\, (2\pi)^3} \Big[ b_{+\vec{k}} \,\,e^{-\Omega_k x^0 + i\vec{k}\cdot\vec{x}}    \\
    &    \quad\quad\quad\quad\quad\quad\quad\quad\quad\quad\quad\quad\quad\quad  + b_{-\vec{k}} \,\,e^{\Omega_k x^0 + i\vec{k}\cdot\vec{x}} \Big]
    \,,
\label{eq:tach_phi}
\ee
and $\omega_k = \sqrt{|\vec{k}|^2 - |m_\phi|^2}$ and $\Omega_k = \sqrt{|m_\phi|^2 - |\vec{k}|^2}$.

The key difference between this approach and cases discussed in the literature, is that indefinite metric is imposed for both stable ($|\vec{k}|>|m_\phi|$) and unstable modes ($|\vec{k}|<|m_\phi|$).
The following commutation relations hold:
\be
\label{eq:CCR}
    &[a_{+\,\vec{k}}, a^\dagger_{+\,\vec{l}}\,] = +(2\pi)^3\, (2\omega_k) \, \delta^3(\vec{k}-\vec{l})\,, \\
    &[a_{-\,\vec{k}}, a^\dagger_{-\,\vec{l}}\,] = -(2\pi)^3\, (2\omega_k) \, \delta^3(\vec{k}-\vec{l})\,, \\
    &[b_{-\,\vec{k}}, b^\dagger_{+\,\vec{l}}\,]= -i (2\pi)^3\, (2\Omega_k) \,\delta^3(\vec{k}-\vec{l})\,, \\
    &[b_{+\,\vec{k}}, b^\dagger_{-\,\vec{l}}\,]= i (2\pi)^3 \,(2\Omega_k)\, \delta^3(\vec{k}-\vec{l})
\,,
\ee
and other commutators vanish, in particular, $[b_{+ \,k},b^\dagger_{+ \,l}]=0$.

We define vacuum as the state normalized to unity that is annihilated by all operators entering $\phi(x)$, \ie, $a_{\pm\,\vec{k}} |0\rangle = b_{\pm\,\vec{k}} |0\rangle = 0$ for all real\footnote{Individual unstable plane waves can acquire complex spatial momenta under Lorentz boosts. Consequently, the real-momentum mode expansion does not by itself define the action of finite boosts on the chosen Fock space; the required domain analysis lies beyond our scope.} three-momenta $\vec{k}$. Then, $\phi(x)|0\rangle = 0$ for all $x$.
As a result, the single particle states that correspond to imaginary energies have zero norm (which is mandatory for a Hamiltonian self-adjoint with respect to the indefinite inner product with complex eigenvalues).
In fact, since the total tachyon field is given by $\chi(x)$, its single particle states of the form $(a^\dagger_{+\,k}+a^\dagger_{-\,k}) |0\rangle$ for stable modes and $(b^\dagger_{+\,k}+b^\dagger_{-\,k}) |0\rangle$  for unstable ones, have zero norm; however, they do not establish a fakeon projection out of the asymptotic state space.

By direct calculation, one verifies \cref{eq:Com} and shows that the time-ordered two point function of $\chi$ is given by the Wheeler propagator ($i\, \Delta_W$).
This shows that expectation values of fields constructed from $\chi$ lead to manifestly LI commutator and Green's functions, and that the CCRs are LI--see \cref{eq:CCR_symplectic,eq:Com}.
Covariance of the \textit{field algebra} does not by itself establish that Lorentz transformations can be implemented by operators preserving the indefinite inner product in Fock-space representation. 
Note that the parametrization $a\sim\sqrt{\omega_k}\,q+i\,p/\sqrt{\omega_k}$ is singular at $\omega_k=0$, and there are boosts that change the sign of the energy, which means that $\phi_>(x)$ cannot be a Lorentz scalar~\cite{Kamoi:1971qb,Paczos:2023mof}.

On the other hand, invariance of the state $|0\rangle$ can be established at formal level by constructing Poincare group generators, and showing they annihilate $|0\rangle$.
Indeed, consider the following generators of time and space translations:
\be
    H&= \int_{|\vec{k}|>|m_\phi|} \frac{d^3\vec{k}}{2(2\pi)^3}\, \left( a^\dagger_{+k} \,a_{+k} + a^\dagger_{-k} \,a_{-k} \right) \\
    &-  \int_{|\vec{k}|<|m_\phi|} \frac{d^3\vec{k}}{2(2\pi)^3}\, \left(b^\dagger_{-k}\, b_{+k} + b^\dagger_{+k}\, b_{-k} \right)
    \,,
\ee
\be
    P^j &= \int_{|\vec{k}|>|m_\phi|} \frac{d^3\vec{k}}{2\omega_k(2\pi)^3}\, k^j \left( a^\dagger_{+k}\, a_{+k} + a^\dagger_{-k}\, a_{-k} \right)  \\
    & -i \int_{|\vec{k}|<|m_\phi|} \frac{d^3\vec{k}}{2\Omega_k(2\pi)^3}\, k^j \left(b^\dagger_{-k}\, b_{+k} - b^\dagger_{+k} \,b_{-k} \right)
    \,,
\ee
where due to the $|\vec{k}|<|m_\phi|$ modes, the Hamiltonian and momentum operators do not have uniform form like for the $m^2>0$ case.
This ansatz satisfies
\be
\label{eq:EoM}
    [P^\mu, \phi(x)] = -i\, \partial^\mu \phi(x)
    \,,
\ee
which explicitly reads ($j=1,2,3$),
\be
    [H, \phi(x)] = -i\, \partial_0 \phi(x)
    \,\,\, \mathrm{and}\,\,\, 
    [P^j, \phi(x)] = i\, \partial_j \phi(x)
\,.
\ee
Since  $P^\mu \,|0\rangle=0$, the vacuum is translation invariant.
On the other hand, spectrum of $H$ is not positive definite—it contains not only positive and negative real eigenvalues, but also imaginary ones. 
Therefore, $|0\rangle$ is an eigenstate with zero four-momentum, but it is not ground state—there are infinitely many zero-energy eigenstates as well as lower energy states.

Moreover, we provide formal expressions for the generators of the homogeneous Lorentz group, assuming integration by parts and dropping the surface terms. 
For rotations and boosts, they read as follows for the stable modes:
\be
    M_{ij,\,>} =& i \int_{|\vec{k}|>|m_\phi|} \frac{d^3\vec{k}}{2\omega_k(2\pi)^3} \Big[ a^\dagger_{+k}\, (k_i \partial_{k_j} - k_j \partial_{k_i})\, a_{+k}  \\
     &- a^\dagger_{-k}\, (k_i \partial_{k_j} - k_j \partial_{k_i})\, a_{-k} \Big]
\,,
\ee
\be
    M_{0i,\,>} =& i \int_{|\vec{k}|>|m_\phi|} \frac{d^3\vec{k}}{2\omega_k(2\pi)^3} \Big[ a^\dagger_{+k}\, (\omega_k \partial_{k_i})\, a_{+k} \\
    &- a^\dagger_{-k}\, (\omega_k \partial_{k_i})\, a_{-k} \Big]
\,.
\ee
Analogous expressions hold for the $|\vec{k}|<|m_\phi|$ modes, 
\be 
    M_{ij,<} =& \int_{|\vec{k}|<|m_\phi|} \frac{d^3\vec{k}}{2\Omega_k(2\pi)^3} \Big[ b_{-k}^\dagger \,(k_i \partial_{k_j} - k_j \partial_{k_i})\, b_{+k} \\
        &- b_{+k}^\dagger\, (k_i \partial_{k_j} - k_j \partial_{k_i})\, b_{-k} \Big]
\,,
\ee
\be
    M_{0i,<} =& -i\int_{|\vec{k}|<|m_\phi|} \frac{d^3\vec{k}}{2\Omega_k(2\pi)^3}
    \Big[
        b_{-\vec{k}}^\dagger \,(\Omega_k \partial_{k_i})\, b_{+\vec{k}} \\
        &+ b_{+\vec{k}}^\dagger \,(\Omega_k \partial_{k_i})\, b_{-\vec{k}}
    \Big]
\,.
\ee

\section{Tachyon-bradyon interactions: scalar background and weak equivalence principle violation}
\label{sec:nonrel_potential}
\begin{figure}[tb]
    \includegraphics[scale=0.24]{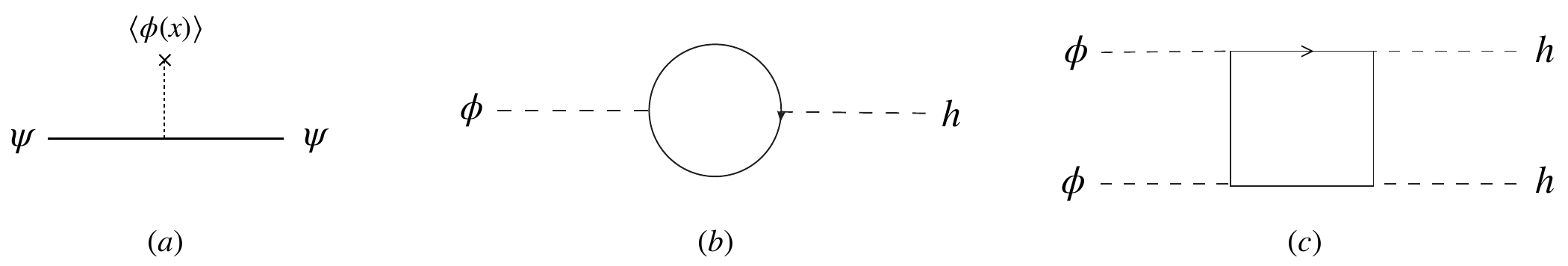}
    \caption{
            Interactions of the virtual tachyon $\phi$: (a) exchange between fermions inducing a long-range oscillating potential; (b) mixing with the Higgs boson; (c) mixed quartic interaction $\phi^2 h^2$.
        }
    \label{fig:tachyon_Higgs_mixing}
\end{figure}
In this section, we show that interactions of a virtual tachyon with ordinary matter generate a matter-dependent scalar background and can lead to violations of the weak equivalence principle in the observable sector.
Here we impose the PVP prescription rather than the time-ordered kernel of the preceding canonical construction, since we are motivated by diagrammatic discussions of fakeons~\cite{Anselmi:2021hab}.
Moreover, the Wheeler propagator has exponentially growing temporal tails, so its ordinary convolution with an eternal static source does not converge; while the static field equation admits stationary particular solutions, they are unprotected from generic perturbations.

Let us first clarify the role of the tachyon mass squared parameter $m_\phi^2$, which is a \textit{bare} parameter that has physical meaning only in free theory.
Once interactions are included, one must distinguish the bare parameter from the pole of the resummed propagator. For bradyon, a pole can be parametrized as $p^2=\mu^2$, with $\mu=m-i\Gamma/2$~\cite{Willenbrock:2022smq}.
A tachyonic pole has no rest-frame decay interpretation, so we discuss its position $\mu_\phi^2$ in the $p^2$ plane directly.

\begin{figure}[tb]
    \includegraphics[scale=0.249]{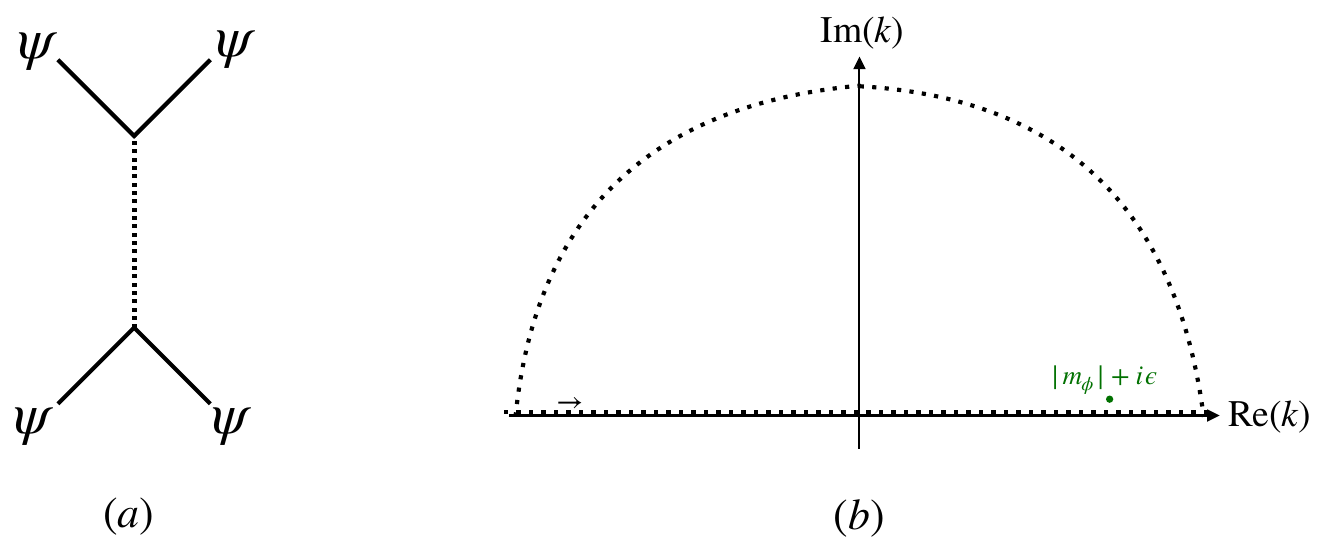}
    \caption{
            (a) Elastic electron scattering mediated by virtual tachyon exchange. (b) The integration contour used to evaluate the non-relativistic potential.
        }
    \label{fig:Yuk}
\end{figure}
For a stable bradyon, $\mu=m$, while $\Gamma \neq 0$ is obtained only if a decay channel becomes kinematically open.
Since the $S$ matrix is unitary (and the same holds for fakeons~\cite{Anselmi:2021hab}), $\Gamma$ can be also calculated by invoking the optical theorem, $\Gamma(q^2) \propto \mathrm{Im}(\mathcal{M})$.
Therefore, for a tachyon to have non-zero Im$(\mu^2)$, it would need to decay.
However, since the tachyon is kinematically forbidden to decay into two bradyons, the imaginary part of its self-energy vanishes (we show the result for virtual scalars),
\begin{align}
\label{eq:M}
    &\mathrm{Im}(\mathcal{M}) = \nonumber\\
    &-\pi\int_0^1 dx \, \theta\left[p^2 x(1-x)-m_0^2(1-x)-m_1^2 x \right] = 0 
    \,, 
\end{align}
where $p^2<0$ and the masses squared of the particles in the loop are positive.
Even tachyon self-interactions (which can modify the value of $m^2$ but have to keep it real) or decays into other tachyons are not viable, since such states would also need to be purely virtual—then, $\mathrm{Im}(\mathcal{M})=0$ also needs to hold.
This shows \textit{stability} of tachyons against particle decays, and thus $\mu_\phi^2$ cannot have non-zero imaginary part; it can only change its value due to renormalization.
However, the field itself clearly is \textit{unstable}, since by equations of motion it wants to roll down the potential.

Moreover, once tachyon interacts with SM fields, it 1) mixes with the Higgs and 2) receives contribution to its effective mass squared parameter after the Higgs acquires a non-zero vev—see panels (b) and (c) in \cref{fig:tachyon_Higgs_mixing}, respectively.
Therefore, interacting tachyon may be turned into bradyon by developing non-zero vacuum expectation value (for the case of positive quartic self-interactions), or equivalently by resumming tadpole diagrams~\cite{Kip:2024uha}, as well as by radiative corrections.
Since we are interested in an interacting tachyon, we assume further that such dynamics does not take place.

Let us now discuss the consequences of introducing Yukawa interactions between a tachyon ($\phi$) and electrons ($\psi$) by examining elastic scattering of the latter, mediated by exchange of virtual latter state—see the panel (a) in \cref{fig:Yuk} for illustration (the $u$ channel is not shown).
Using \cref{eq:FP} in momentum space and the contour shown in the panel (b) in \cref{fig:Yuk}, we obtain the following non-relativistic potential resulting from tachyon exchange:
\begin{align}
\label{eq:pot}
    V_T(|\vec{r}|) &= -g^2\,\mathrm{Re} \int \frac{d^3\vec{k}}{(2\pi)^3} \frac{e^{i\, \vec{k} \cdot \vec{r}}}{|\vec{k}|^2+m_\phi^2 -i\epsilon} \nonumber \\
    &= -g^2\,\mathrm{Re}\,\, \frac{e^{i \sqrt{|m_\phi|^2+i\epsilon} r}}{4\pi r} =\frac{-g^2}{4\pi\,r} \cos(|m_\phi|\,r)
\,,
\end{align}
which gives a long-range oscillating force.
We note that in Ref.~\cite{Perivolaropoulos:2016ucs,Antoniou:2017mhs}, bounds on such oscillating potentials have been discussed.
In the notation of~\cite{Antoniou:2017mhs}, the bound is $\alpha_O \lesssim 10^7$ at $\lambda \simeq  35\,\mu$m.
For the electron coupling considered here, the scalar charge of a neutral atom is approximately $g\,Z$.
Assuming the interaction is unscreened and taking $Z/A\simeq1/2$ gives $|g|\lesssim 2\times10^{-15}$ at $|m_\phi|\simeq0.035\,\mathrm{eV}$.

The static potential given by \cref{eq:pot} also illustrates a distinctive feature of the tachyonic dispersion relation. 
Since the potential is \textit{static}, it corresponds to $\partial \phi/\partial t=0$. Thus, in Fourier space, the energy must vanish, $\omega=0$, which corresponds to the states with momenta $|\vec{k}|=|m_\phi|$ on the shell.

Moreover, the cosine term introduces a characteristic spatial wavelength $\lambda=2\pi/|m_\phi|$, so the resulting background depends on the positions of the sources. For a finite set of stationary nonrelativistic electron sources, a formal static background is
\be
\label{eq:phi_vev}
    \langle\phi(t,\vec{x})\rangle = g \sum_i \frac{\cos(|m_\phi| |\vec{x}-\vec{x}_i|)}{4\pi \, |\vec{x}-\vec{x}_i|}
\,.
\ee
This background is generally spatially nonuniform, depends on the matter distribution, and modifies the effective electron mass as follows:
\be
    m^{\mathrm{effective}}_e = m_e - g \,\langle\phi(x)\rangle 
\,.
\ee
This leads to violation of the weak equivalence principle due to composition-dependent accelerations originating from gradients of the background (electron number per unit mass depends on composition).

\section{Conclusions}
\label{sec:conclusions}
Tachyonic behavior of classical or quantum fields is conventionally viewed as an instability in the long wavelength regime, as indicated by the presence of complex energies in the infrared mode decomposition. 
We have shown that the canonical formalism gives the Wheeler propagator, which retains exponentially growing timelike tails, and the tachyonic PVP cannot be the ordinary time-ordered two-point correlation function of a local Hermitian field with the canonical commutator.
This excludes the direct tachyonic extension of the canonical free-fakeon realization.
We also discussed consequences of using PVP for describing tachyon interactions with ordinary matter.
For the electron coupling, we showed that it generates a long-range oscillatory force and composition-dependent accelerations violating the weak equivalence principle, and we provided a bound on the coupling constant.

In conjunction with our previous work~\cite{Jodlowski:2024rut}, we have established that formulation of covariant QFT of on-shell or fakeon tachyon is not viable within canonical QFT formalism.
While we have not excluded formulation of tachyon QFT that would abandon one of the standard axioms, such as relativistic covariance, locality, and unitarity, any such attempt would not only face stringent constraints\footnote{For example, the recent work~\cite{Hertzberg:2025pph} has put a limit on tachyon mass within a non-local field theory of tachyons. Since our starting point is relativistic covariance, and keeping only the $|\vec{k}|>|m_\phi|$ modes is LI only on-shell, such theories lie beyond our scope.} but also question about naturalness of such an approach.
Finally, our work provides further support to lack of viability of asymptotic freedom in higher-derivative quantum gravity~\cite{Avramidi:1985ki}, and shows that constructions employing tachyons within Carrollian physics face obstacles if extended to Minkowski space.

\begin{acknowledgments}
This work was supported by the National Natural Science Foundation of China (NNSFC) under grants No.~12335005, No.~12575118, and the Special funds for postdoctoral overseas recruitment, Ministry of Education of China. 
This work was also supported by the Institute for Basic Science under the project code, IBS-R018-D1. 

We thank Andrzej Dragan, Jan Dereziński, and Stanisław Mrówczyński for discussions, comments, and useful suggestions.
We thank Damiano Anselmi, Ryszard Horodecki, Matthew Lake, and Marco Piva for correspondence.
\end{acknowledgments}

\bibliography{bibliography}

\end{document}